\documentclass[prd,amsmath,amssymb,floatfix,nofootinbib]{revtex4}
\usepackage{graphicx,dcolumn,booktabs}
\usepackage{amsmath}
\usepackage{amsfonts,color}
\usepackage{amssymb}
\usepackage{times}
\usepackage{amsmath, amssymb}
\usepackage[dvipdfm,  
            colorlinks, 
            citecolor=blue
            ]{hyperref}

\begin{document}
\title{The FSI contribution to the observed $B_s$ decays into $K^+K^-$ and $\pi^+K^-$}

\author{Xu-Hao Yuan$^{1}$}\email{segoat@mail.nankai.edu.cn}
\author{Hong-Wei Ke$^{2}$}\email{khw020056@hotmail.com, Corresponding author}
\author{Xiang Liu$^3$}\email{xiangliu@lzu.edu.cn}
\author{Xue-Qian Li$^{4}$}\email{lixq@nankai.edu.cn}
\affiliation{
  $^{1}$Center for High Energy Physics, Tsinghua University, Beijing 100084,
  China\\
  $^{2}$School of Science, Tianjin University, Tianjin 300072,
  China\\
  $^3$School of Physical Science and Technology, Lanzhou University, Lanzhou 730000, China\\
  $^{4}$School of Physics, Nankai University, Tianjin 300071,
  China }

\begin{abstract}
\noindent Because at the tree level  $B_s\rightarrow K^+K^-$ is
Cabibbo-triple suppressed, so its branching ratio should be
smaller than that of $B_s\to \pi^+ K^-$. The measurements present
a reversed ratio as $R=\mathcal B(B_s\rightarrow
\pi^+K^-)/\mathcal B(B_s\rightarrow K^+ K^-)\sim{4.9/33}$.
Therefore, It has been suggested that the transition $B_s\to
K^+K^-$ is dominated by the penguin mechanism, which is
proportional to $V_{cb}V^*_{cs}$.  In this work, we show that an
extra contribution from the final state interaction (FSI) to
$B_s\to K^+K^-$ via sequential processes $B_s\to D^{(*)}_{(s)}
\bar D_s^{(*)}\to K^+K^-$ is also substantial and should be
superposed on the penguin contribution. Indeed, taking into
account of the FSI effects, the theoretical prediction on $R$ is
well consistent with the data.
\end{abstract}

\maketitle

\section{Introduction}\label{sec-1}

It is generally believed that the hierarchy of the
Cabibbo-Kabayashi-Maskawa (CKM) matrix elements determines the
magnitudes of weak decays, and off-diagonal matrix elements would
bring up an order suppression. For example, the ratio of $\mathcal
B(B_s\rightarrow D_s^+\pi^-)/\mathcal B(B_s\rightarrow D_s^\pm
K^\mp)=3.2\times10^{-3}/3.0\times10^{-4}$ has been measured
{\cite{Abulencia:2006aa,Louvot:2008sc}}, and it is roughly
determined by the ratio of the CKM matrix elements
$V_{ud}/V_{us}$. Thus one usually categorizes the weak decays
according to their CKM structures as the Cabibbo-favored;
Cabibbo-suppressed and even the Cabibbo-double or
triple-suppressed. However, the newly observed modes
$B_s\rightarrow K^+K^-$ and $B_s\rightarrow \pi^+K^- $ obviously
do not follow the rule, namely at first look, the branching ratio
of $B_s\rightarrow K^+K^-$ should be smaller than that of
$B_s\rightarrow \pi^+ K^-$ by $|V_{us}/V_{ud}|^2$, by contraries,
the datum is $\mathcal{R}=\mathcal B(B_s\rightarrow
\pi^+K^-)/\mathcal B(B_s\rightarrow
K^+K^-)={4.9}\times10^{-6}/3.3\times10^{-5}$
\cite{PDG10,Abulencia:2006psa,Aaltonen:2008hg}. If only the tree
diagrams are taken into account, this reversion would compose an
``anomaly". Descotes-Genon et al. \cite{DescotesGenon:2006wc}
carefully analyzed the transitions of $B_s\to K^0\bar K^0$ and
$K^+K^-$ through flavor symmetries and QCD factorization, and
pointed a potential conflict between the QCD prediction on $B_s\to
K^+K^-$ and data. By analyzing the transition mechanism, Cheng and
Chua \cite{Cheng} determined that the main contribution to $B_s\to
K^+K^-$ comes from the penguin diagram, which is proportional to
$V_{cb}V^*_{cs}$.  Ali et al. also calculated the branching ratios
of $B_s\to \pi^+K^- $ and $B_s\to K^+K^-$ in terms of the
Perturbative QCD (PQCD) to LO and NLO \cite{Ali:2007ff,Liu:2008rz}
and the authors of Ref. \cite{Williamson:2006hb} did calculations
in SCET. Their results roughly were consistent with the data
available then, so the ``conflict" seemed resolved. Even though
the theoretical uncertainties in all the calculations are not
fully controlled, it is noted that the obtained central values are
not sufficiently large to make up the data. It implies that there
must be some mechanisms to remarkably enhance the branching ratio
of $B_s\rightarrow K^+K^-$.

Looking at the central values they obtained and the resultant ratio
of $\mathcal{B}(B_s\to \pi^+K^-)/\mathcal{B}(B_s\to K^+K^-)$, one
can find that the calculated $\mathcal{B}(B_s\to \pi^+K^-)$ is close
to the data, however, the central values of $\mathcal{B}(B_s\to
K^+K^-)$ calculated in various approaches are smaller than the newly
measured data \cite{Abulencia:2006psa,Aaltonen:2008hg}.

Based on this observation, we suggest that the final state
interaction (FSI) in $B_s$ decays may greatly enhance the branching
ratio of $B_s\to K^+K^-$ but not much for $B_s\to \pi^+K^-$. In fact, in the
energy regions of $b$-quark and $c$-quark most such anomalies can be
naturally explained by considering the role of FSI. For example, a simple quark diagram-analysis
tells that the branching ratio of $D^0\to K^0\bar K^0$ is almost
zero, but its measured value is comparable with that of $D^0\to
K^+K^-${,} which is large. This can be understood by considering the
sequential process $D^0\to K^+K^-\to K^0\bar K^0$ and the later step
is a hadronic scattering \cite{Dai:1999cs}.

The {Particle Data Group (PDG)} \cite{PDG10} tells us that the
$D_s^{(*)+}D_s^{(*)-}$ are the dominant hadronic decay modes of
$B_s$, therefore it implies that the sequential decays $B_s\to
D_s^{(*)+}D_s^{(*)-}\to K^+K^-$ would compose an important
contribution to the observed $B_s\to K^+K^-$. The first step of
$B_s\to D_s^{(*)+}D_s^{(*)-}$ is only suppressed by $V_{bc}$ and
the process does not suffer form a color-suppression. Moreover it
is also isospin-conserved mode, even though the weak interaction
does not require isospin conservation, it still may be more
favorable than the isospin violated ones. Thus, one can understand
why $B_s\to D_s^{(*)+}D_s^{(*)-}$ is dominant. Then let us look at
the next step. The FSI is a hadronic scattering process, which is
completely governed by strong interaction, so that the isospin
must be conserved. The isospin of $D_s^{(*)\pm}$ is zero, thus,
the final state of the scattering is required to be zero. The
isospin of $K$-mesons is 1/2, thus the $K\bar{K}$-states can be
either isospin 0 or 1, therefore the inelastic scattering
$D_s^{(*)+}D_s^{(*)-}\to K^+K^-$ is allowed. By contraries,
isospin of pion is 1, thus, the system of $\pi^+K^-$ does not
contain an isospin 0 component, thus the scattering
$D_s^{(*)+}D_s^{(*)-}\to \pi^+K^-$ is forbidden. Definitely, one
can expect a substantial contribution from $B_s\to
D_s^{(*)+}D_s^{(*)-}\to K^+K^-$, which can much enhance the
branching ratio of $B_s\to K^+K^-$ in comparison with $B_s\to
\pi^+K^-$.

It is worth pointing out that on the other hand, the decay
$B_s\rightarrow \pi^+K^-$ also receives a contribution from the FSI
via $B_s\rightarrow D^{(*)+}D_s^{(*)-}\rightarrow \pi^+K^-$.
However, the first step process $B_s\rightarrow D^{(*)+}D_s^{(*)-}$
is Cabibbo-double suppressed by $V_{cb}V_{cd}^*$, so is much less
than $B_s\rightarrow D_s^{(*)+}D_s^{(*)-}$, therefore the FSI does
not contribute much to the branching ratio of $B_s\rightarrow
\pi^+K^-$.

Notice that, the direct process $B_s\to K^+K^-$ via penguin diagrams
and the sequential process with FSI $B_s\rightarrow
D_s^{(*)}D_s^{(*)}\rightarrow K^+K^-$ have the same initial and
final states, moreover their amplitudes are of the same order of
magnitude, so the two contributions interfere. In fact, their
contributions are not directly experimentally, but theoretically
distinguishable. Moreover, the strong scattering would have a real
and an imaginary parts (see below, for the calculations of the
triangle diagrams), thus a phase is resulted. In most calculations
of the FSI effects, only the absorptive (imaginary) part is kept,
the reason is that one may argue that the absorptive part might be
dominant or at most the dispersive and absorptive parts have close
magnitudes. In that case, the absorptive part is imaginary while the
penguin contribution is real, so that the two contributions can be
added up at the rate level (amplitude square). Our final results
indicate that  while taking into account the new contribution, the
theoretical predictions are indeed close to the new data.

In this work, we calculate the amplitudes of the decay channels of
$B_s\rightarrow D_s^{(*)}(D^{(*)})\bar{D}_s^{(*)}$ and the direct decay
channels of $B_s\rightarrow K^+(\pi^+)K^-$ by the factorization
approach \cite{Matinyan:1998cb,Ablikim:2002ep}. Then we use the
effective SU(4) Lagrangian \cite{Haglin:1999xs,Lin:1999ad} to
determine the vector-pseudoscalar-pseudoscalar and
vector-vector-pseudoscalar vertices for calculating the FSI
amplitude of $D_s^{(*)}(D^{(*)})\bar{D}_s^{(*)}\rightarrow K^+(\pi^+)K^-$.  The
paper is organized as follows, after the introduction, we formulate
the weak decays and the re-scattering processes in section II, our
numerical results are presented along with all necessary inputs in
section III. The last section is devoted to discussion and
conclusion.

\section{The theoretical evaluations of the branching ratios}
\subsection{Weak decays in factorization approach}\label{sec-2}
Because the measurements on the decay widths of $B_s\to
D_s^{(*)+}D_s^{(*)-}$ and {$B_s\to D^{(*)+}D_s^{(*)-} (B_s\to
D^{(*)-}D_s^{(*)+})$} are not accurate yet, as for most of the
above channels there are only upper bounds, so that we are going
to directly calculate the transition amplitudes based on the quark
diagrams. Even though this strategy might bring up certain
theoretical uncertainties, it does not break our qualitative
conclusion at all.

For calculating the transition amplitudes of $B_s\rightarrow
D_s^{(*)+}(D^{(*)+})D_s^{(*)-}$, one needs to employ the effective
hamiltonian at the quark level. With the operator product
expansion (OPE), the effective Hamiltonian was explicitly
presented in Ref. \cite{Buchalla:1995vs}. At the tree level, from
the effective Lagrangian one can notice that $B_s\to K^+K^-$ is
suppressed by $V_{ub}V^*_{us}$ i.e triple-Cabibbo suppressed,
comparing with $B_s\to \pi^+K^-$, which is double-Cabibbo
suppressed by $V_{ub}$. Therefore Cheng and Chua decided that the
direct transition $B_s\to K^+K^-$ is obviously dominated by the
penguin diagram whose CKM structure is $V_{cb}V^*_{cs}$. The
transition $B_s\to D^{(*)+}_s D^{(*)-}$ is also double-Cabibbo
suppressed. Instead, $B_s\to D^{(*)+}_s D^{(*)-}_s$ is
proportional to $V_{cb}V^*_{cs}$ and it is a tree process. Now let
us compare the sequential processes $B_s\to D^{(*)}\bar
D_s^{(*)}\to K^+K^-$ with the penguin contribution. The two
reactions are of the same  CKM structure, but the penguin
undergoes a loop suppression about $\alpha_s/\pi\sim 0.06$,
whereas the strong scattering $D_s^{(*)}\bar D_s^{(*)}\to
\sum_iX_i$ where $X_i$ stands as any possible final states allowed
by symmetry and energy-momentum conservation. $K^+K^-$ is only one
of the possible channels, and its probability is proportional to
$\langle D_s^{(*)}\bar D_s^{(*)}|H_{eff}|K\bar{K}\rangle$, which
is what we are going to calculate in this work. This is a
suppression factor because the total probability to all channels
is 1.  Therefor, roughly, we notice that the penguin is
loop-suppressed and the sequential process is also suppressed by
the probability, thus the two modes compete and may have a similar
order of magnitude. Concretely, we need to calculate them. The
explicit calculation on the penguin contribution can be found in
\cite{Ali:2007ff,Liu:2008rz,Cheng}, thus we will use their numbers
and only consider the contribution from the sequential processes.

Applying the
effective hamiltonian at the quark level to the hadron states, the
hadronic matrix elements can be parameterized as \cite{Cheng:2003sm}:
\begin{eqnarray}\label{decay matrix}
 \langle0|J_\mu|P(p_1)\rangle&=&-if_Pp_{1\mu},\nonumber\\
 \langle0|J_\mu|V(p_1,\epsilon)\rangle&=&f_V\epsilon_\mu m_V,\nonumber
 \end{eqnarray}
 \begin{eqnarray}
 \langle P(p_2)|J_\mu|B_s(p)\rangle&=&\Big[P_\mu-{m_{B_s}^2-m_P^2\over q^2}q_\mu\Big]F_1(q^2)+{m_{B_s}^2-m_P^2\over q^2}q_\mu F_0(q^2),\nonumber
  \end{eqnarray}
 \begin{eqnarray}
 \langle V(p_2,\epsilon)|J_\mu|B_s(p)\rangle&=&{i\epsilon^\nu\over m_{B_s}+m_V}
 \Big\{i\epsilon_{\mu\nu\alpha\beta} P^\alpha q^\beta A_V(q^2)+(m_{B_s}+m_V)^2g_{\mu\nu}A_1(q^2)-{P_\mu P_\nu\over m_{B_s}+m_V}A_2(q^2)\nonumber\\
  & &-2m_V(m_{B_s}+m_V){P_\nu q_\mu\over q^2}[A_3(q^2)-A_0(q^2)]\Big\},
\end{eqnarray}
where $J_\mu=\bar q_1\gamma_\mu(1-\gamma_5)q_2$,
$P_\mu=(p_1+p_2)_\mu$ and $q_\mu=(p_1-p_2)_\mu$.

With Eq. (\ref{decay
matrix}), we write down the amplitudes of $B_s\rightarrow D_s^{(*)+}
D_s^{(*)-}$:
\begin{subequations}
\begin{eqnarray}\label{amplitude Bs2DsDs}
 \mathcal A(B_s(p)\rightarrow D_s^+(p_1) D_s^-(p_2))&=&-{iG_F\over\sqrt2}V_{cb}V^*_{cs}a_1f_{D_s}(m_{B_s}^2-m_{D_s}^2)F_0^{B_sD_s}(p_1^2),\\
 \mathcal A(B_s(p)\rightarrow D^{*+}_s(p_1) D^{*-}_s(p_2))&=&{G_F\over\sqrt2}V_{cb}V^*_{cs}a_1f_{D_s^*}m_{D_s^*}{ig^{\mu\nu}\over m_{B_s}+
 m_{D_s^*}}\Big\{i\epsilon_{\mu\nu\alpha\beta}(p_1+2p_2)^\alpha p_1^\beta A_V^{B_sD_s^*}(p_1^2)+(m_{B_s}\nonumber\\
 & &+m_{D_s^*})^2g_{\mu\nu}A_1^{B_sD_s^*}(p_1^2)-(p_1+2p_2)_\mu(p_1+2p_2)_\nu A_2^{B_sD_s^*}(p_1^2)-{(p_1+2p_2)_\mu p_{1\nu}\over p_1^2}\Big[(m_{B_s}\nonumber\\
 & &+m_{D^*_s})^2A_1^{B_sD_s^*}(p_1^2)-(m_{B_s}^2-m_{D^*_s}^2)A_2^{B_sD_s^*}(p_1^2)-2m_{D^*_s}(m_{B_s}+m_{D^*_s})A_0^{B_sD_s^*}(p_1^2)\Big]\Big\},\nonumber\\
\end{eqnarray}
and the amplitudes of $B_s\rightarrow D^{(*)+}D_s^{(*)-}$ read as
\begin{eqnarray}
 \mathcal A(B_s(p)\rightarrow D^+(p_1)D_s^-(p_2))&=&-{iG_F\over\sqrt2}V_{cb}V^*_{cd}a_1f_D(m_{B_s}^2-m_{D_s}^2)F_0^{B_sD_s}(p_1^2),\\
 \mathcal A(B_s(p)\rightarrow D^{*+}(p_1)D_s^{*-}(p_2))&=&{G_F\over\sqrt2}V_{cb}V^*_{cd}a_1f_{D^*}m_{D^*}{ig^{\mu\nu}\over m_{B_s}
 +m_{D_s^*}}\Big\{i\epsilon_{\mu\nu\alpha\beta}(p_1+2p_2)^\alpha p_1^\beta A_V^{B_sD_s^*}(p_1^2)+(m_{B_s}\nonumber\\
 & &+m_{D_s^*})^2g_{\mu\nu}A_1^{B_sD_s^*}(p_1^2)-(p_1+2p_2)_\mu(p_1+2p_2)_\nu A_2^{B_sD_s^*}(p_1^2)-{(p_1+2p_2)_\mu p_{1\nu}\over p_1^2}\Big[(m_{B_s}\nonumber\\
 & &+m_{D^*_s})^2A_1^{B_sD_s^*}(p_1^2)-(m_{B_s}^2-m_{D^*_s}^2)A_2^{B_sD_s^*}(p_1^2)-2m_{D^*_s}(m_{B_s}+m_{D^*_s})A_0^{B_sD_s^*}(p_1^2)\Big]\Big\}, \nonumber\\
\end{eqnarray}
the amplitude of  $\pi^+K^-$ at the tree level is:
\begin{eqnarray}
 \mathcal A^\mathrm{direct}(B_s(p)\rightarrow \pi^+(p_1)K^-(p_2))&=&-{iG_F\over\sqrt2}V_{ub}V^*_{ud}a_1f_\pi(m_{B_s}^2-m_K^2)F_0^{B_sK}(p_1^2),
\end{eqnarray}
\end{subequations}
where $a_1$ is a proper combination of the Wilson coefficients in
the effective hamiltonian
\cite{Buchalla:1995vs,Cheng:1986an,Li:1988hr}. $F_0(q^2)$ and
$A_{V,1,2}(q^2)$ are the form factors to be detrmined. In this work,
due to lack of accurate data, we use the form factors obtained by
fitting the data of the decays  of $B$ meson. It is a reasonable
approximation because the processes $B_s\rightarrow D_s^*$ and
$B\rightarrow D^*$ have the same topological structure, and  the
flavor $SU(3)$ symmetry for light quarks ($u$, $d$ and $s$) would
lead to  the same form factors, namely the difference between the
form factors for different light-quark flavors would be proportional
to an $SU(3)$ breaking, which is small for the effective vertices as
well known. At least such small difference would not overtake the
errors caused by experimental measurements and theoretical
uncertainties for evaluating the non-perturbtive effects.

Moreover, a symmetry analysis indicates that the sequential process
$B_s\to D_s^{*+} D_s^-(D_s^+ D^{*-}_s)\to K^+K^-$ is forbidden by
angular-momentum conservation.

Taking the three-parameter form, the form factors are written as
\cite{Cheng:2003sm}:
\begin{eqnarray}
 F(q^2)={F(0)\over1-a{q^2\over m_B^2}+b{\left(q^2\over m_B^2\right)}^2},
\end{eqnarray}
where $a$, $b$ and $F(0)$ are the three parameters and their values
are listed in Table \ref{formfactor}.
\renewcommand{\arraystretch}{1.5}
\begin{table}[htb]
\begin{tabular}{cccc|cccc}
  \hline\hline
  $F$ & $F(0)$ & $a$ & $b$ & $F$ & $F(0)$ & $a$ & $b$ \\
  \midrule[1pt]
  $F_0^{BD}$ & 0.67 & 0.65 & 0.00 & $F_0^{B\pi}$ & 0.25 & 0.84 & 0.10  \\
  $V^{BD^*}$ & 0.75 & 1.29 & 0.45 & $A_0^{BD^*}$ & 0.64 & 1.30 & 0.31 \\
  $A_1^{BD^*}$ & 0.63 & 0.65 & 0.02 & $A_2^{BD^*}$ & 0.61 & 1.14 & 0.52 \\
  \hline\hline
\end{tabular}
\caption{The parameters given in Ref. \cite{Cheng:2003sm}.}\label{formfactor}
\end{table}

\subsection{Evaluation of FSI effects }
\begin{figure}[htb]
  \centering
  \includegraphics[width=12cm]{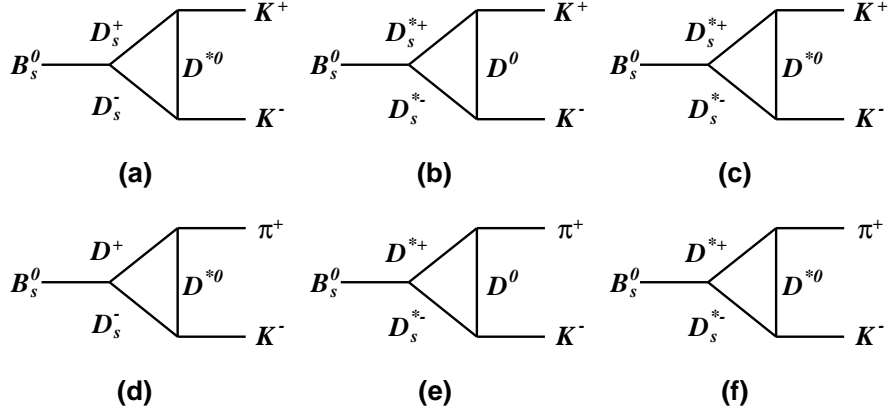}
  \caption{The QCD one-particle exchange process in $D_s^{(*)+}D_s^{(*)-}\rightarrow K^+K^-$.}\label{fig-DsDs2KK}
\end{figure}
Now let us turn to evaluate the long-distance effects at hadron
level. For  the effective vertices, the flavor-$SU(4)$ symmetry is
assumed. The coupling of pseudoscalar and vector meson is
\begin{eqnarray}
 \mathcal L_0=\mathrm{Tr}(\partial_\mu \phi^\dagger\partial^\mu \phi)-{1\over2}\mathrm{Tr}(F^\dagger_{\mu\nu}F^{\mu\nu}),
\end{eqnarray}
where $F^{\mu\nu}=\partial^\mu V^\nu-\partial^\nu V^\mu$, and
$\phi$ and $V$ represent the $4\times 4$ pseudoscalar and vector
meson matrices in $SU(4)$, {respectively}:
\begin{eqnarray}
 \phi&=&{1\over\sqrt2}
 \begin{pmatrix}
  {\pi^0\over\sqrt2}+{\eta\over\sqrt6}+{\eta_c\over\sqrt12} & \pi^+ & K^+ & \bar D^0 \\
  \pi^- & -{\pi^0\over\sqrt2}+{\eta\over\sqrt6}+{\eta_c\over\sqrt12} & K^0 & D^- \\
  K^- & \bar K^0 & -\sqrt{2\over3}\eta+{\eta_c\over\sqrt12} & D^-_s \\
  D^0 & D^+ & D^+_s & -{3\eta_c\over\sqrt12},
 \end{pmatrix},
 \end{eqnarray}
 \begin{eqnarray}
 V&=&{1\over\sqrt2}
 \begin{pmatrix}
  {\rho^0\over\sqrt2}+{\omega'\over\sqrt6}+{J/\psi\over\sqrt12} & \rho^+ & K^{*+} & \bar D^{*0} \\
  \rho^- & -{\rho^0\over\sqrt2}+{\omega'\over\sqrt6}+{J/\psi\over\sqrt12} & K^{*0} & D^{*-} \\
  K^{*-} & \bar K^{*0} & -\sqrt{2\over3}\omega'+{J/\psi\over\sqrt12} & D^{*-}_s \\
  D^{*0} & D^{*+} & D^{*+}_s & -{3J/\psi\over\sqrt12}.
 \end{pmatrix}.
\end{eqnarray}
For the gauge invariance,  covariant derivatives  replace the
regular ones:
\begin{eqnarray}
\begin{aligned}
 \partial_\mu\phi&\rightarrow\mathcal D_\mu\phi=\partial_\mu\phi-{ig\over2}[V_\mu,\phi],\\
 F_{\mu\nu}&\rightarrow\partial_\mu V_\nu-\partial_\nu V_\mu-{ig\over2}[V_\mu,V_\nu].
\end{aligned}
\end{eqnarray}
Now we are ready to write down the relevant terms in the
pseudoscalar-vector coupling:
\begin{eqnarray}\label{lagrangian_int}
 \mathcal L=\mathcal L_0+ig\mathrm{Tr}(\partial^\mu\phi[\phi,V^\nu])+g'\epsilon_{\alpha\beta\mu\nu}\partial^\alpha V^\beta\partial^\mu V^\nu\phi+\cdots,
\end{eqnarray}
where the third term on the right-hand side of
Eq. (\ref{lagrangian_int}) is involved additionally as a
vector-vector-pseudoscalar coupling \cite{Haglin:2000ar}. In the
process that $D_s^{(*)+}D_s^{(*)-}$ re-scatter into $K^+K^-$, the
corresponding Lagrangian is:
\begin{eqnarray}\label{lagrangian_KDD}
\begin{aligned}
 &\mathcal L_{KD_sD^*}=ig_{KD_sD^*}\Big\{D^{-*}_\mu\Big[\partial^\mu K^+D_s^--K^+\partial^\mu D_s^-\Big]+\bar D^{0*}\Big[K^-\partial^\mu D_s^+-\partial^\mu K^-D_s^+\Big]\Big\},\\
 &\mathcal L_{KD_s^*D}=ig_{KD_s^*D}\Big\{D^{-*}_{s\mu}\Big[K^+\partial^\mu D^0-\partial^\mu K^+D^0\Big]+D^{+*}_{s\mu}\Big[\partial^\mu K^-\bar D^0-K^-\partial^\mu\bar D^0\Big]\Big\},\\
 &\mathcal L_{KD_s^*D^*}=g_{KD_s^*D^*}\epsilon^{\alpha\beta\mu\nu}\Big[\partial_\alpha\bar D^{0*}_\beta\partial_\mu D^{+*}_{s\nu}K^-+\partial_\alpha D^{-*}_{s\beta}\partial_\mu D^{0*}_\nu K^+\Big],
\end{aligned}
\end{eqnarray}
and the effective vertices for $\pi DD^*$ and $\pi D^*D^*$ are
written as \cite{Haglin:1999xs,Liu:2007qs}.
\begin{eqnarray}\label{lagrangian_piDD}
\begin{aligned}
 &\mathcal L_{\pi DD^*}=\frac{i}{2}g_{\pi DD^*}(\bar D \tau_iD^{*\mu}\partial_\mu\pi_i-\partial^\mu\bar D \tau_iD^{*}_\mu\pi_i-H.c.)
 ,\\
 &\mathcal L_{\pi D^*D^*}=-g_{\pi D^*D^*}\varepsilon^{\mu\nu\alpha\beta}\partial^\mu\bar D^*_\nu\pi\partial_\alpha D^{*\beta}
\end{aligned}
\end{eqnarray}
In the Eq. (\ref{lagrangian_KDD}), the values of the coupling
constants should be obtained by fitting the experiment data.
However, it is noticed that in previous literature  those coupling
constants are still not well fixed yet (see Table \ref{coupling
constant}), so in this work, we use an average value for each
coupling constant, and the uncertainty is considered as a
systematical error, which might be attributed to our input
parameters. For example, there are no available data for determining
the coupling constants $g_{KD_sD^*}$ and $g_{KD_s^*D}$, we are going
to fix them based on the $SU(4)$ symmetry, which tells us that:
$g_{KD_sD^*}=g_{KD_s^*D}=\sqrt2g_{\rho\pi\pi}={\sqrt{3}}g_{\psi
DD}/2$, and by fitting the data,  $g_{\rho\pi\pi}=8.8$ and $g_{\psi
DD}=7.9$ are set (see Table \ref{coupling constant}). We may have
two different values for $g_{KD_sD^*}$ and $g_{KD_s^*D}$: $6.0$ and
$9.4$, but by our strategy we adopt an average value for
$g_{KD_sD^*}$ and $g_{KD_s^*D}$ as:
$g_{KD_sD^*}=g_{KD_s^*D}=7.7\pm1.7$. We use the same method to get
the values for other relevant coupling constants as long as there
are no data to directly fix them, and retain the errors. Then we
have:
\begin{eqnarray}
\begin{aligned}
 &g_{KD_sD^*}=g_{KD_s^*D}=\sqrt2g_{\rho\pi\pi}={\sqrt{3}\over2}g_{\psi DD}=7.7\pm0.9,\\
 &{g_{KD_s^*D^*}={1\over\sqrt3}g_{\psi D^*D}=(2.40\pm0.02)}\,\mathrm{GeV}^{-1},\\
 &g_{\pi D^*D^*}=(9.0\pm0.1)\,\mathrm{GeV}^{-1},\quad
{ g_{\pi DD^*}=8.8.}
\end{aligned}
\end{eqnarray}

The authors of Ref. \cite{Casalbuoni:1996pg} gave a simple
relation between $g_{\pi D^*D^*}$ and  $g_{\pi DD^*}$ as $g_{\pi
D^*D^*}= 1.4 g_{\pi DD^*}$ which can also be  used to determine
$g_{\pi DD^*}$ or $g_{\pi D^*D^*}$ from each other.

To evaluate the FSI effects, we calculate the absorptive part (see
above arguments) of the one-particle-exchange triangle diagrams.
Fig. \ref{fig-DsDs2KK} shows the diagrams of $B_s\rightarrow
D_s^{(*)+}D_s^{(*)-}\rightarrow K^+K^-$ by exchanging $D^{0(*)}$. For
calculating the absorptive part of the triangle, with the Cutkosky
cutting rule \cite{Cheng:2004ru,Shifman:1978bx}, the related
amplitudes are:
\begin{subequations}\label{amplitude Bs2DsDs2KK}
\begin{eqnarray}
 & &\mathcal A_1\Big[B_s(p)\rightarrow D_s^+(p_1)D_s^-(p_2)\rightarrow K^+(p_3)K^-(p_4)\Big]\nonumber\\
 &&={1\over2}\int d\tilde p_1d\tilde p_2(2\pi)^4\delta(p-p_1-p_2)\mathcal A\Big[B_s\rightarrow D_s^+(p_1)D_s^-(p_2)\Big](-g_{KD_sD^*})\Big[-i(p_1+p_3)_\mu\Big](-g_{KD_sD^*})\Big[-i(p_2+p_4)_\nu\Big]\nonumber\\
 & &\quad\times\Big[-g_{\mu\nu}+{q_\mu q_\nu\over m_{D^{0*}}^2}\Big]{i\over q^2-m_{D^{0*}}^2}\mathcal F^2(q^2,m_{D^{0*}}^2),
  \end{eqnarray}
 \begin{eqnarray}
 & &\mathcal A_2\Big[B_s(p)\rightarrow D_s^*(p_1)D_s^*(p_2)\rightarrow K^+(p_3)K^-(p_4)\Big]\nonumber\\
 &&={1\over2}\int d\tilde p_1d\tilde p_2(2\pi)^4\delta(p-p_1-p_2)\mathcal A\Big[B_s\rightarrow D_s^{+*}(p_1)D_s^{-*}(p_2)\Big](-g_{KD_s^*D})\Big[-i(q+p_3)_\xi\Big](-g_{KD_s^*D})\Big[-i(q-p_4)_\sigma\Big]\nonumber\\
 & &\quad\times\Big[-g_{\mu\xi}+{p_{1\mu}p_{1\xi}\over m_{D_s^*}^2}\Big]\Big[-g_{\nu\sigma}+{p_{2\nu}p_{2\sigma}\over m_{D_s^*}^2}\Big]{i\over q^2-m_{D^0}^2}\mathcal F^2(q^2,m_{D^0}^2),
  \end{eqnarray}
 \begin{eqnarray}
 & &\mathcal A_3\Big[B_s(p)\rightarrow D_s^{*+}(p_1)D_s^{*-}(p_2)\rightarrow K^+(p_3)K^-(p_4)\Big]\nonumber\\
 &&={1\over2}\int d\tilde p_1d\tilde p_2(2\pi)^4\delta(p-p_1-p_2)\mathcal A\Big[B_s\rightarrow D_s^{+*}(p_1)
 D_s^{-*}(p_2)\Big](ig_{KD_s^*D^*})(-ip_{1\sigma})(-iq_\alpha)\epsilon^{\sigma\omega\alpha\beta}(ig_{KD_s^*D^*})(-ip_{2\xi})(iq_\kappa)\nonumber\\
 & &\quad\times\epsilon^{\xi\lambda\kappa\rho}\Big[-g_{\beta\rho}+{q_\beta q_\rho\over m_{D^{0*}}^2}\Big]
 \Big[-g_{\mu\omega}+{p_{1\mu}p_{1\omega}\over m_{D_s^*}^2}\Big]\Big[-g_{\nu\lambda}+{p_{2\nu}p_{2\lambda}\over m_{D_s^*}^2}\Big]{i\over q^2-m_{D^{0*}}^2}\mathcal F^2(q^2,m_{D^{*0}}^2),
\end{eqnarray}
instead, for the decay channel $B_s\rightarrow
D^{(*)+}D_s^{(*)-}\rightarrow \pi^+K^-$, we have:
\begin{eqnarray}
 & &\mathcal A_1\Big[B_s(p)\rightarrow D^+(p_1)D_s^-(p_2)\rightarrow \pi^+(p_3)K^-(p_4)\Big]\nonumber\\
 &&={1\over2}\int d\tilde p_1d\tilde p_2(2\pi)^4\delta(p-p_1-p_2)\mathcal A\Big[B_s\rightarrow D^+(p_1)D_s^-(p_2)\Big](-g_{\pi DD^*})\Big[-i(p_1+p_3)_\mu\Big](-g_{KD_sD^*})\Big[-i(p_2+p_4)_\nu\Big]\nonumber\\
 & &\quad\times\Big[-g_{\mu\nu}+{q_\mu q_\nu\over m_{D^{0*}}^2}\Big]{i\over q^2-m_{D^{0*}}^2}\mathcal F^2(q^2,m_{D^{0*}}^2),\\
 & &\mathcal A_2\Big[B_s(p)\rightarrow D^{*+}(p_1)D_s^{*-}(p_2)\rightarrow \pi^+(p_3)K^-(p_4)\Big]\nonumber\\
 &&={1\over2}\int d\tilde p_1d\tilde p_2(2\pi)^4\delta(p-p_1-p_2)\mathcal A\Big[B_s\rightarrow D^{+*}(p_1)D_s^{-*}(p_2)\Big](-g_{\pi DD^*})\Big[-i(q+p_3)_\xi\Big](-g_{KD_s^*D})\Big[-i(q-p_4)_\sigma\Big]\nonumber\\
 & &\quad\times\Big[-g_{\mu\xi}+{p_{1\mu}p_{1\xi}\over m_{D^*}^2}\Big]\Big[-g_{\nu\sigma}+{p_{2\nu p_{2\sigma}}\over m_{D_s^*}^2}\Big]{i\over q^2-m_{D^0}^2}\mathcal F^2(q^2,m_{D^0}^2),
 \end{eqnarray}
 \begin{eqnarray}
 & &\mathcal A_3\Big[B_s(p)\rightarrow D^{*+}(p_1)D_s^{*-}(p_2)\rightarrow \pi^+(p_3)K^-(p_4)\Big]\nonumber\\
 &&={1\over2}\int d\tilde p_1d\tilde p_2(2\pi)^4\delta(p-p_1-p_2)\mathcal A\Big[B_s\rightarrow D^{+*}(p_1)
 D_s^{-*}(p_2)\Big](ig_{\pi D^*D^*})(-ip_{1\sigma})(-iq_\alpha)\epsilon^{\sigma\omega\alpha\beta}(ig_{KD_s^*D^*})(-ip_{2\xi})(iq_\kappa)\nonumber\\
 & &\quad\times\epsilon^{\xi\lambda\kappa\rho}\Big[-g_{\beta\rho}+{q_\beta q_\rho\over m_{D^{0*}}^2}\Big]\Big[-g_{\mu\omega}
 +{p_{1\mu}p_{1\omega}\over m_{D^*}^2}\Big]\Big[-g_{\nu\lambda}+{p_{2\nu}p_{2\lambda}\over m_{D_s^*}^2}\Big]{i\over q^2-m_{D^{0*}}^2}\mathcal F^2(q^2,m_{D^{*0}}^2),
\end{eqnarray}
\end{subequations}
where ${d\tilde p}=dp^3/((2\pi)^32E)$, $q$ is the momentum of the
exchanged $D^{0(*)}$ meson and a {dipole} form factor $\mathcal
F(q^2,m^2)=(\Lambda^2-m^2)^2/(\Lambda^2-q^2)^2$ with
$\Lambda=m+\beta\Lambda_{QCD}$ \footnote{Here, $m$ denotes the
mass of the exchanged meson and $\beta$ is a phenomenological
parameter which is set to 1 in our calculation. $\Lambda_{QCD}$ is
taken as 220 MeV.} is introduced to compensate its off-shell
effect \cite{Liu:2007qs}.
\renewcommand{\arraystretch}{1.5}
\begin{table}[htb]
\begin{tabular}{cc|cc}
  \hline\hline
  $g$ & value & $g$ & value \\
  \midrule[1pt]
$g_{D^*D\psi}$ & $4.2\mathrm{GeV}^{-1}$ \cite{Liu:2007qs}, ${\frac{7.7 }{m_D}}$\,{GeV}$^{-1}$ \cite{Haglin:1999xs} & $g_{DD\psi}$ & 7.9 \cite{Liu:2007qs}, 8.0 \cite{Deandrea:2003pv}\\
   $g_{\pi DD^*}$ & 8.8 \cite{Haglin:1999xs} & $g_{\pi D^*D^*}$ & $8.9\, \mathrm{GeV}^{-1}$ \cite{Liu:2007qs}, 9.1\,$\mathrm{GeV}^{-1}$ \cite{Oh:2000qr}\\
    \hline\hline
\end{tabular}
\caption{The values of the coupling constants involved in our
work.}\label{coupling constant}
\end{table}

\section{Numerical result}

With the amplitudes given in Eq. (\ref{amplitude Bs2DsDs2KK}), we
can easily calculate the decay width of the sequential processes
$B_s\rightarrow D_s^{(*)}D_s^{(*)}\rightarrow K^+K^-$. We take the
relevant parameters from PDG \cite{PDG10} as: $m_{B_s}=5.366\,
\mathrm{GeV}$, $m_{D_s}=1.968\, \mathrm{GeV}$, $m_{D^*_s}=2.112\,
\mathrm{GeV}$, $m_{D^0}=1.864\, \mathrm{GeV}$, $m_{D^{0*}}=2.007\,
\mathrm{GeV}$, $m_{D^\pm}=1.869\, \mathrm{GeV}$,
$m_{D^{*\pm}}=2.01\, \mathrm{GeV}$; and the value of the weak
decay constants, such as $f_{\pi}\; f_K\; f_D\; f_{D_s}$ can be
found in Ref. \cite{Azizi:2008ty}; $a_1=1.14$ is set in our
numerical computations \cite{Ablikim:2002ep}. The total amplitudes
are
\begin{eqnarray}
 \Big|\mathcal A(B_s\rightarrow K^+K^-)\Big|&=&\Big|\mathcal A^\mathrm{direct}(B_s\rightarrow K^+K^-)+\mathcal A_1(B_s\rightarrow D_s^+D_s^-\rightarrow K^+K^-)+\mathcal A_2(B_s\rightarrow D_s^{*+}D_s^{*-}\rightarrow K^+K^-)\nonumber\\
 & &+\mathcal A_3(B_s\rightarrow D_s^{*+}D_s^{*-}\rightarrow K^+K^-)\Big|=(4.1\pm0.8)\times10^{-8},
  \end{eqnarray}
 \begin{eqnarray}
 \Big|\mathcal A(B_s\rightarrow \pi^+ K^-)\Big|&=&\Big|\mathcal A^\mathrm{direct}(B_s\rightarrow \pi^+ K^-)+\mathcal A_1(B_s\rightarrow D^+D_s^-\rightarrow \pi^+ K^-)+\mathcal A_2(B_s\rightarrow D^{*+}D_s^{*-}\rightarrow \pi^+ K^-)\nonumber\\
 & &+\mathcal A_3(B_s\rightarrow D^{*+}D_s^{*-}\rightarrow \pi^+ K^-)\Big|=30.4^{+0.2}_{-0.1}\times10^{-9},
\end{eqnarray}
where $A^\mathrm{direct}(B_s\rightarrow K^+K^-)$ is the penguin
contribution \cite{Cheng}. So, the branching ratios are:
\begin{eqnarray}\label{Br}
 \mathcal B(B_s\rightarrow K^+K^-)&=&{\Gamma_{B_s\rightarrow K^+K^-}\over\Gamma_\mathrm{tot}}=
 {1\over32\pi^2\Gamma_\mathrm{tot}}{\Big|p_K\Big|\over m_{B_s}^2}\Big|\mathcal A_\mathrm{tot}(B_s\rightarrow KK)\Big|^2d\Omega\nonumber=13.6^{+6.3}_{-5.1}\times10^{-6},\nonumber\\
 \mathcal B(B_s\rightarrow \pi^+K^-)&=&{\Gamma_{B_s\rightarrow \pi K}\over\Gamma_\mathrm{tot}}=
 {1\over32\pi^2\Gamma_\mathrm{tot}}{\Big|p_K\Big|\over m_{B_s}^2}\Big|\mathcal A_\mathrm{tot}(B_s\rightarrow K\pi)\Big|^2d\Omega\nonumber=7.8^{+0.1}_{-0.1}\times10^{-6},\nonumber\\
\end{eqnarray}
where the errors are systematical, originating from the uncertainty
of the coupling constants.
\renewcommand{\arraystretch}{1.5}
\begin{table}[htb]
\caption{Various approaches predictions on  $\mathcal{B}(B_s\to
K^+K^-)$ (in units of $10^{-6}$).} and the ratio
$\mathcal{R}=\mathcal B(B_s\rightarrow \pi^+K^-)/\mathcal
B(B_s\rightarrow K^+K^-)$\label{Tab:t7}
\begin{ruledtabular}
\begin{tabular}{cccccccccc}
 &QCDF \cite{Cheng} & pQCD (LO) \cite{Ali:2007ff} & pQCD (NLO) \cite{Liu:2008rz}& SCET \cite{Williamson:2006hb} &FSI&FSI+pQCD (NLO)& Experiment \cite{Abulencia:2006psa,Aaltonen:2008hg}\\\midrule[1pt]
$\mathcal{B}$&$ 25.2^{+12.7+12.5}_{-7.2-9.1}$&
$13.6^{+8.6}_{-5.2}$&$15.6^{+5.1}_{-3.9}$  &
$18.2\pm6.7\pm1.1\pm0.5$ & ${13.6^{+6.3}_{-5.1}}$&
${29.2^{+8.1}_{-6.5}}$&$33\pm9$\\\hline
$\mathcal{R}$\footnote{Since the main contribution for the
transition $B_s\rightarrow \pi^+K^-$ comes from the tree diagram
and all predictions made in various models on this channel  are
close to each other, we use the data $\mathcal{B}(B_s\rightarrow
\pi^+K^-)=4.9\times 10^{-6}$ as input in our calculations.}
&0.194&0.360&0.314&0.269&0.360&0.167&0.148
\end{tabular}
\end{ruledtabular}
\end{table}

In Table \ref{Tab:t7} we list the theoretical predictions on
$\mathcal{B}(B_s\to K^+K^-)$ in various approaches. It is noted that
most of the predicted cental values  are lower than the newly
measured value  \cite{Cheng,Ali:2007ff,Liu:2008rz,Williamson:2006hb}
as long as the contribution from FSI is not included. We add  the
contributions from FSI to that calculated in pQCD (NLO), then one
can find that the  resultant central value is consistent with data
\cite{Beringer:1900zz}: $\mathcal B(B_s\rightarrow
K^+K^-)=(3.3\pm0.9)\times10^{-5}$ within $1\sigma$.

The deviations of our theoretical prediction from the data might
come from the loophole in our calculation. As indicated above we
only consider the absorptive part of the hadronic triangle. Indeed
the dispersive part may also make substantial contributions
\cite{Liu:2006dq}. As suggested in literature, the dispersive
contribution should be smaller than the absorptive one (it is
consistent with the general principle of the quantum field theory),
or at most has the same magnitude as that of the absorptive part. If
the contribution of the dispersive part is indeed of the same order
as the absorptive part, then taking it into account, we may have a
result, which is even closer to the data.

In general, even though we cannot precisely re-produce the
experimental data, we can confirm ourselves that the FSI is
important and non-negligible  for understanding the
$\mathcal{B}(B_s\rightarrow K^+K^-)>\mathcal{B}(B_s\rightarrow
\pi^+K^-)$ ``conflict".

\section{Conclusion and Discussion}

As expected, the LHCb is extensively aiming on study of the
B-physics, especially to look for some "anomalies" in experiments,
which need high statistics and precise measurements. For $B_s$'s
charmless non-leptonic two-body decay, the early MC studies show
that, nearly 37K $B_s\rightarrow K^+K^-$ signals will be seen at $2\
fb^{-1}$ integrated luminosity \cite{Rademacker:2007zza}. On the
other hand, $314\pm 27$ $B_s\rightarrow \pi^+K^-$ signals were observed
based on the 2010 data with its integrated luminosity $0.35fb^{-1}$
\cite{Aaij:2012qe}. We believe that the statistics of the $B_s$
decay into $K^+K^-$ and $\pi^+K^-$ is sufficient to draw a definite decision
about their branching ratios. The latest result reported by the LHCb
collaboration shows that the branching ratios have been measured as
\cite{:2012as}:
\begin{eqnarray}
 {\mathcal{B}(H_b\rightarrow F)\over\mathcal{B}(H_b'\rightarrow F')}={f_{H'_b}
 \over f_{H_b}}{N(H_b\rightarrow F)\over N(H'_b\rightarrow F')}{\epsilon_\mathrm{rec}(H'_b\rightarrow F')
 \over\epsilon_\mathrm{rec}(H_b\rightarrow F)}{\epsilon_\mathrm{PID}(F')\over\epsilon_\mathrm{PID}(F)},
\end{eqnarray}
where the $f_{H^{(')}_b}$ is the possibility of $b$ quark
hadronizing into hadron $H$, $N$ is the observed number of signals
for certain decay modes, $\epsilon_\mathrm{rec}$ is the efficiency
of the reconstruction excluding the particle identification (PID)
cuts and $\epsilon_\mathrm{PID}$ is just the efficiency of PID cuts.

It is noted that in Table III, the experimental data are taken from
Refs. \cite{Abulencia:2006psa,Aaltonen:2008hg}, but the 2012 data of PDG indicate that the branching
ratio of $B_s\to K^+K^-$ is $(2.64\pm 0.28)\times 10 ^{-5}$
\cite{Beringer:1900zz}, which is smaller than the data in Ref. \cite{Abulencia:2006psa,Aaltonen:2008hg}. Moreover, the LHCb
collaboration reports that with the $0.37fb^{-1}$ 2011 data, the
branching ratios of $B_s\rightarrow K^+K^-$ and $B_s\rightarrow
\pi^+K^-$ are experimentally determined as:
\begin{eqnarray}
\begin{aligned}
 \mathcal B(B_s\rightarrow K^+K^-)&=(23.0\pm0.7\pm2.3)\times 10^{-6},\\
 \mathcal B(B_s\rightarrow \pi^+K^-)&=(5.4\pm0.4\pm0.6)\times 10^{-6},
\end{aligned}
\end{eqnarray}
where the former uncertainties are statistical and the later one is
systematical, which include the uncertainties of PID calibration,
final state radiation with soft gamma, signal shape used for
fitting, and the impact of the background: the additional three-body
background\footnote{Miss or misidentify the pion or kaon in final
state.}, the combinatorial background and the cross-feed
background\footnote{The uncertainties from the distribution of
signal in data and simulation.}. Since $B_s\rightarrow K^+K^-$ and
$B_s\rightarrow D^{(*)+}_s\bar D^{(*)-}_s\rightarrow K^+K^-$ have
the same final states, it means we cannot single out the
contributions of FSI by simply measuring the cross section in
experiments. Even though our theoretical prediction on
$\mathcal{B}(B_s\to K^+K^-)$ presented in Table \ref{Tab:t7} is
slightly above the value of the LHCb measurements, it is still
consistent with the data within the experimental error tolerance.
Therefore we are expecting  more precise measurements in the future.

Cheng and Chua suggested that the decay $B_s\to K^+K^-$ is
dominated by the penguin diagram, which does not suffer from large
CKM suppression \cite {Cheng}. In that scenario the ``conflict"
$\mathcal{B}(B_s\to K^+K^-)> \mathcal{B}(B_s\to \pi^+K^-)$ could
be partly explained. In our work, we show that the FSI definitely
makes an important contribution because $B_s\to D_s^{(*)}\bar
D_s^{(*)}$ are dominant decay modes of $B_s$ as confirmed by the
data and the scattering $D_s^{(*)+} D_s^{(*)-}\to K^+K^-$ is
allowed by all the symmetry requirements.  Therefore, we might
consider that the penguin contribution and the FSI effects are in
parallel to contribute to the decay $B_s\to K^+K^-$.\footnote{In
this work we do not consider the penguin contribution, but only
that of the FSI effects to the decay. We acknowledge the possible
contribution from the penguin diagram as well, further study
involving both of them and moreover their interference will be
made in our later works.} Actually, there are several
phenomenological parameters in the model for evaluating the FSI
effects, which were obtained by fitting the earlier data of heavy
flavor processes. One thing can be sure that the FSI should exist
and contribute to the process $B_s\to K^+K^-$, but how significant
it would be is determined by both requirement of experimental
measurements and theoretical estimate. If the data on
$\mathcal{B}(B_s\to K^+K^-)$ are indeed going down, the FSI
effects for $B_s\to K^+K^-$ would be less significant, by that
situation, one can further restrict the involved phenomenological
parameters at this energy region. On other aspect, the errors of
the measurements are still too large to make a definite conclusion
yet, so we are expecting more precise measurements not only from
LHCb, but also the future super-B factory.

Definitely the FSI effects also play roles in other similar decay
channels, such as $B^\pm\rightarrow K^\pm\omega$. Therefore careful
studies on such modes with the same theoretical framework would be
helpful for more accurately evaluating the FSI effects. The PDG
indicates that, $\mathcal B(B^\pm\rightarrow K^\pm\omega)/\mathcal
B(B^\pm\rightarrow
\pi^\pm\omega)=(6.7\times10^{-6})/(6.9\times10^{-6})$, so one would
expect that $\mathcal B(B\rightarrow K\omega)$ is also enhanced by
the FSI. Moreover, a measurement on the polarization of $\omega$
might be helpful to gain more information about the role of the FSI
mechanism, and it will be studied in our coming work.

From the experimental aspect, as the FSI is a scattering fully
governed by the strong interaction, its proper time is too short to
be measured in the LHCb detector. Although the LHCb can well
reconstruct the events $B_s\rightarrow K^+(\pi^+)K^-$ and identify
the $\pi^+K^-$ purely with the PID cuts \cite{Karbach:2012sw}, it is
impossible to distinguish between the direct decays $B_s\to
K^+(\pi^+)K^-$ and the sequential ones. Considering the good ability
of LHCb for tracking charged particles, we believe that,
$B^\pm\rightarrow K(\pi)^\pm\omega$ is another good channel to study
FSI. The  decay rates of $\mathcal B(B\rightarrow \bar D^0
D^{(*)+})$ have been well measured as $\sim1.8\pm0.2\times10^{-2}$,
and we can make Mont-Carlo simulations on the re-scattering of $\bar
D^0 D_s^{(*)+}$ into $K(\pi)^+\omega$ to complete the theoretical
scenario. The abnormal ratio of the two channels $\mathcal
B(B^\pm\rightarrow K^\pm\omega)/\mathcal B(B^\pm\rightarrow
\pi^\pm\omega)=(6.7\times10^{-6})/(6.9\times10^{-6})$
\cite{Beringer:1900zz} can also be explained as the FSI contribution
as well\footnote{$\mathcal B(B\rightarrow
K(\pi)\omega)=6.7(6.9)\times10^{-6}$, at the same time, the Cabibbo
suppression determines that the branching ratio of $K\omega$ final
state should be smaller.}.
With $B^\pm\rightarrow DK(\pi)^\pm$ measurement \cite{Aaij:2012kz}
and considering the efficiency, about 0.8K $B^+\rightarrow
K(\pi)^+\omega$ can been seen in the 2011 database and it implies
that the FSI contribution to the polarization of the vector meson
$\omega$ might be distinguished from that of the direct decay.
Comparing more accurate data with our theoretical calculation, one
can expect to gain more knowledge on  the FSI effects, for example
how to determine the parameters in the dipole form factor etc.

One object can be recommended for pining down the role of the FSI
effects. As well known that the direct CP violation is proportional
to $\sin(\alpha_1 -\alpha_2)\cdot\sin(\phi_1-\phi_2)$, where the
subscripts 1 and 2 refer to two different routes and $\alpha$,
$\phi$ are the weak and strong phases respectively. Therefore, if a
direct CP violation exists in $B_s$ decays, there must at least be
two different routes to the final states which possess different
weak and strong phases. In fact,  the penguin and tree diagrams have
different strong and weak phases. The weak phase of the tree diagram
is coming from $V_{ub}V^*_{us}$ and does not posses a strong phase.
Instead, the weak phase of the penguin is from $V_{cb}V^*_{cs}$ and
the strong phase is due to existence of the absorptive part of the
loop. In our scenario the sequential processes $B_s(\bar B_s)\to
D_s^{(*)}\bar D_s^{(*)}\to K^+K^-$ which have the same weak phase
$V_{cb}V^*_{cs}$, but different strong phase from the penguin.
Because of the extra contribution, the amplitude would become
$$M=M^{tree}e^{i\alpha_1}+M^{penguin}e^{i(\alpha_2+\phi_1)}+M^{seq}e^{i(\alpha_2+\phi_2)},$$
where the last term is the FSI
contribution.

Thus the interference between the tree diagram with a sum of the
penguin and sequential processes would be different from the
situation where the tree diagram only interferes with the penguin.
Therefore by measuring the CP violation of  $\mathcal{B}(B_s\to K^+K^-)-\mathcal{B}(\bar
B_s\to K^+K^-)$, one can distinguish between the penguin
contributions and the FSI effects. But since the experimental errors
and theoretical uncertainties are not well controlled so far, we
cannot make a more definite prediction on the CP violation yet, we
would wait for more precise data to be available and then continue
our theoretical computations.

Moreover, according to the above arguments, there does not exist a
tree diagram for $B_s\to K^0\bar K^0$. Consequently, the direct
transition $B_s\to K^0\bar K^0$ is uniquely determined by the
penguin diagram and FSI effects. Because the tree contribution to
$B_s\to K^+K^-$ is very small, the rate of $B_s\to K^0\bar K^0$ must
be close to that of $B_s\to K^+ K^-$. However the interference
between penguin (neglecting the Cabibbo-suppressed penguin diagrams
where $t$-quark and $u$-quark are intermediate agents) and FSI does not
lead to a direct CP violation at all, because they have the same
weak phase. This can be confirmed by future experiments.

\section*{Acknowledgments}

We would like to thank Ming-Xing Luo and Cai-Dian Lu for useful
discussion. This project is supported by the National Natural
Science Foundation of China (NSFC) under Contract Nos. 11075079,
11005079, 11222547, 11175073 and 11035006, the Ministry of
Education of China (SRFDF under Grant Nos. 20100032120065 and
20120211110002, FANEDD under Grant No. 200924, NCET under Grant
No. NCET-10-0442, the Fundamental Research Funds for the Central
Universities), and the Fok Ying-Tong Education Foundation (No.
131006).

\vfil

\end{document}